# Nanometer-Scale Materials Contrast Imaging with a Near-Field Microwave Microscope

Atif Imtiaz[1] and Steven M. Anlage
Center for Superconductivity Research, Department of Physics, University of Maryland, College Park, MD 20742-4111
John D. Barry and John Melngailis
Institute for Research in Electronics and Applied Physics, University of Maryland, College Park, MD  20742-3511

We report topography-free materials contrast imaging on a nano-fabricated Boron-doped Silicon sample measured with a Near-field Scanning Microwave Microscope over a broad frequency range. The Boron doping was performed using the Focus Ion Beam technique on a Silicon wafer with nominal resistivity of 61 $\Omega$.cm. A topography-free doped region varies in sheet resistance from 1000$\Omega/\square$ to about 400k$\Omega/\square$ within a lateral distance of 4μm. The qualitative spatial-resolution in sheet resistance imaging contrast is no worse than 100 nm as estimated from the frequency shift signal.

---

[1] Present email: atif@boulder.nist.gov

As CMOS (Complimentary Metal-Oxide-Semiconductor) technology reaches the sub-45 nm node, and operating frequencies of integrated circuits reach into the microwave regime, there is a strong need to quantitatively measure local materials properties at microwave frequencies with nanometer spatial-resolution [1,2]. For example, there is a need to develop metrological tools to characterize ultra-shallow-doped films for sub-micron semiconductor devices [3], which requires local measurement of materials properties such as sheet-resistance ($R_x$). Similarly, metrological tools are required to understand the basic physics of carbon nano-tubes and nano-wires before they become useful for electronics [1]. To achieve such goals, Near-field Scanning Microwave Microscopes (NSMM) have been developed as quantitative metrology tools [4,5], and with such microscopes, the local $R_x$ has been measured with spatial resolution on the micrometer length scale [6-8]. However, sample topography affects the measured NSMM signals [9] and it is quite challenging to separate the topographic effects from materials properties [7]. Such challenges raise questions regarding the ability of NSMM to be a useful metrology tool for nano-science and nano-technology. In order to demonstrate the NSMM as a useful metrology tool at such length scales, we designed a topography-free sample with varying $R_x$ on the nanometer length scale. This sample, prepared using Focused Ion Beam (FIB) implantation [10], (described in detail elsewhere [11, 12]), has been measured with a high-resolution NSMM which utilizes Scanning Tunneling Microscopy (STM) for distance-following [13] and is capable of measuring $R_x$ variations of thin films [7].

The sample is a 10 μm x 10 μm patch with a sheet resistance gradient ranging from $R_x$ = 30 Ω/□ to 5.5x10$^5$ Ω/□ over a lateral length scale of 10 μm (Fig. 1). However, the topography-free region has values in the range $R_x$ = 10$^3$ Ω/□ to 4x10$^5$ Ω/□, as will be discussed shortly. The STM topography image of the variable-$R_x$ region of the sample is shown in Fig.1(a). Because the doping concentration is constant in the vertical direction we can take the average lateral line cuts of STM topography (the open squares in Fig. 1(b) on the right vertical axis). The calculated $R_x$ profile of the sample as a function of position is also shown in Fig. 1(b) (dashed line on left vertical axis). This $R_x$ profile is calculated using the FIB beam parameters [10] and the knowledge of the implanted Boron depth and profile [11,12]. The region of high Boron concentration (and low $R_x$) is on the left side of the sample. This part of the sample is damaged (labeled so in Fig. 1(b)) because the beam dwell time was very long on this region to dope to a high concentration. As a result there is roughly 10 nm of topography present in the high Boron concentration region. The extreme right hand side of the doped area close to the un-doped Silicon also shows some topography of about 5 nm. In between, there is a lateral region of 4 μm which on average has zero (±1 nm) topography, labeled as 'topography-free' region in Fig. 1(b). This is a good region for maintaining a constant height of the probe above the sample while the NSMM contrast here is expected to be due to the changes in $R_x$ alone.

While maintaining the NSMM probe at constant height of 1 nm above the sample [14], the quality factor (Q) and frequency shift (Δf) images were acquired as shown in Fig 1(c) and 1(d) where the frequency of the experiment is 7.472 GHz. The doped region clearly shows Q and Δf contrast in the topography-free region. The NSMM responds to the variable-$R_x$ region of the sample through both reactive and resistive interactions with the sample. The NSMM also responds to the damaged (strong topography) area. Since the sample has no $R_x$ variation in the vertical direction, one can take lateral line segments at a fixed vertical position from the topography-free area, and average them together to form the line cuts (shown in Fig 2 for Δf and Fig. 3 for Q). The data over the region of damage (as seen on the left side of the images), and the data with topography close to bare Silicon on the right are not shown in these averages. Similar images and analysis were also obtained at microwave frequencies of 1.058 GHz, 3.976 GHz, and 9.602 GHz, with all experiments performed at room temperature.

A lumped element model [4] of the microscope/sample interaction used to understand the data is shown schematically in the inset of Fig. 2. The sample in this case is a thin doped layer with circuit impedance $AR_x$, on top of bulk Silicon, with impedance $Z_{Si}$. $R_x$ is the sheet resistance of the doped layer and $A$ is a geometrical coefficient relating the circuit impedance to the field impedance [4], which arises due to the geometry of the probe and presence of different boundary conditions [15, 16] in the sample. The substrate impedance $Z_{Si}$ includes the resistivity $\rho_{sub}$ of the bulk Silicon ($\rho_{sub}$ = 61 $\Omega$.cm) and the (assumed real) dielectric constant ($\varepsilon_{Si}$ = 11.9). The capacitance between the (STM-tip) inner-conductor of the transmission line resonator and the sample is labeled as $C_x$, and the capacitance between the sample and outer conductor is labeled as $C_{out}$ ($C_{out} \sim$ pF $\gg C_x \sim$ fF).

The expressions [4] to calculate the quality factor and frequency shift from the near-field model include the following parameters. The transmission line resonator (with characteristic impedance $Z_0$ = 50$\Omega$, attenuation $\alpha$, the mode number $n$ and the length $L_{res}$) is terminated by the STM tip (electrical tip), with a characteristic size (called $D_t$) and the sample. The tip and sample together have impedance $Z_{tE}$ which is the sum of the resistive ($R_{tE}$) and the reactive ($X_{tE}$) impedances. Assuming that $|Z_{tE}| \gg Z_0$, the complex reflection coefficient [2] is calculated (for the geometry shown in the inset of Fig. 2) for the resonant condition of the transmission line resonator. This yields predictions for the quality factor Q' and relative frequency shift $\Delta f/f_0$ for the resonator as shown in [4]. The effective dielectric constant of the resonator is $\varepsilon_{eff}$ = 2.5, $L_{res}$ = 1.06 m and attenuation $\alpha$ is frequency dependent (0.0673 Np/m at 1.058 GHz to 0.209 Np/m at 9.602 GHz).

The $R_x$ contrast in the $\Delta f/f_0$ signal saturates below an $R_x$ of about 10 k$\Omega$/□ and above about 400 k$\Omega$/□ for this probe geometry and frequency range, as seen in Fig. 2. For the $R_x$ values in between, the signal shows the largest contrast range. We expect the sensitivity of microscope frequency shift to sheet resistance to be maximized near the condition that the magnitude of the capacitive reactance is equal to the resistive impedance of the sample presented to the probe tip [13]. This is true for all frequencies, as we note that $\omega C_x(AR_x) \cong 1$ roughly in the middle of the $\Delta f/f_0$ contrast range (dots in Fig. 2). For example, for the frequency of 1.058 GHz, this point is near $R_x$ = 75,000 $\Omega$/□ and $\omega C_x(AR_x)$ = 1.2. The Q'/$Q_0$ vs. sheet resistance data (Fig. 3 where $Q_0$ is the quality factor with no sample present) also shows a local minimum where the magnitude of the capacitive reactance is equal to the resistive impedance (shown as black dots) of the sample presented to the probe tip [7, 13].

To fit the data in Figs. 2 & 3, the probe characteristic dimension $D_t$ was fixed to 15 μm (which is twice the typical $R_0$ value when the probe is regarded as a sphere of radius $R_0$ [14]). The other fitting parameters are the tip-to-sample capacitance $C_x$ and the geometrical factor $A$. In Figs. 2 & 3 we show the fit to the $\Delta f/f_0$ and Q'/$Q_0$ data (solid lines), respectively at different frequencies, with the fit parameters shown in Table 1. The fits to $\Delta f/f_0$ vs $R_x$ for all frequencies are excellent, while the same fits also describe the high-$R_x$ Q'/$Q_0$ data quite well. The $C_x$ fit value systematically decreases as the frequency increases because the tip-sample capacitance $C_x$ can be regarded as a series combination of two capacitors; one from the geometry of the tip and the sample ($C_{geometry}$) and the second is the material dependent capacitor ($C_{material}$) [17]. The skin depth of the Silicon substrate is also decreasing as the frequency increases, and as a result the value of $C_{material}$ is effectively increasing. Hence at higher frequencies, $C_{geometry}$ is becoming the more dominant capacitor approaching the value ($\approx$ 1 fF) we expect for tip-sample capacitance [13] calculated from a conducting sphere above an infinite conducting plane model.

Besides the tip-sample capacitance $C_x$, the other fitting parameter is *A* [4], which links the material properties of the sample to $Z_{tE}$, and is dependent on the geometry of the tip [4] and the sample [15,16]. For an ideal coaxial Corbino-geometry contact (with $r_{out}$ = 840μm and $r_{in}$ = 255μm), we expect the value of *A* to be 0.19 [18]. However, *A* can vary if the geometry deviates from the right-cylindrical Corbino-structure, which is the case in our system due to a conical-tip (with embedded sphere at the end [14]), and the geometrical features on the sample (with and without topography). The non-topographic features on the sample include the 10 x 10 μm$^2$ Boron-doped patches and the finite thickness of the Silicon wafer (smaller than both the wavelength and the skin-depth for the measured frequency range). The geometrical features with topography (in the low $R_x$ region) are about 4 x 10 μm$^2$ in size. The presence of topography in this region increases the dissipated power locally [16], affecting primarily the quality factor, as seen in Fig. 3 for $Q`/Q_0$ in the low $R_x$ region (the value of *A* in parentheses in Table I fits the minimum in this region). As the frequency of measurement increases, the value of *A* decreases due to the decreasing skin-depth and associated changes in field configuration.

To conclude, the Boron-doped Silicon sample has a 4 μm x 10 μm region of topography-free variable sheet resistance. This region shows a continuous variation in $R_x$ from values of 1000Ω/□ to 400kΩ/□. As seen in Fig. 2, the frequency shift signal shows a clear contrast in this region. Looking at the data in the region where the magnitude of the capacitive reactance is equal to the resistive impedance of the sample presented to the probe tip ($\omega C_x(AR_x) \cong 1$), two data points that are 100 nm apart on the sample show a clear and distinct contrast in $\Delta f/f_0$. This suggests that the quantitative spatial-resolution in $R_x$ is no worse than 100 nm for the NSMM employed here (consistent with [13]). By designing probes with larger values of the geometry factor *A*, one can increase sensitivity to lower $R_x$ films and materials while maintaining high spatial resolution.

We thank Andrew Schwartz and Vladimir Talanov of Neocera/Solid State Measurements, Inc for insightful discussions. This work has been supported by NSF GOALI (DMR-0201261) and NSF/ECS-0322844.

Table I: The fit parameters ($C_x$, $A$) values for the fits shown as solid lines in Figs 2 and 3 at the corresponding frequencies for fixed probe characteristic dimension $D_t$. The $A$ parameter values in the brackets fit the low-$R_x$ quality factor data minima, and create the fits shown as dashed lines in Fig. 3.

| f(GHz) | $D_t$ (μm) | $C_x$ (fF) | A |
|---|---|---|---|
| **1.058** | 15 | 16.5 | 0.15 (3) |
| **3.976** | 15 | 10.3 | 0.05 (0.65) |
| **7.472** | 15 | 7.7 | 0.05 (0.50) |
| **9.602** | 15 | 5.0 | 0.047 (0.27) |

# Figures

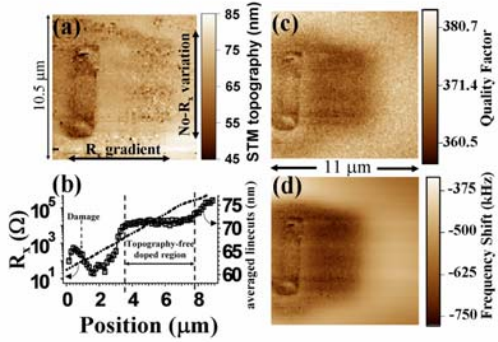

Fig 1: The a) STM topography b) the average of horizontal line cuts taken from the topography image, along with calculated $R_x$ profile, c) Quality Factor and d) Frequency Shift images. The microscope frequency is 7.472 GHz, tunnel current set point for these simultaneous images is 0.5 nA with bias of 1 volt.

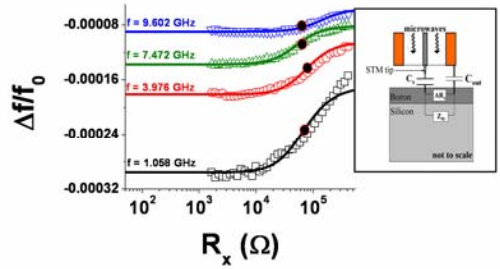

Fig 2: Symbols show the averaged $\Delta f/f_0$ line cuts through Fig. 1(d) for four different frequencies. Also shown are $\Delta f/f_0$ fits to the data, where $f_0$ is the resonant frequency with no sample present. The inset shows the lumped-element model used to fit the data. The fit parameters are given in Table I. The dots on the curve signify points where $\omega C_x(AR_x) = 1$.

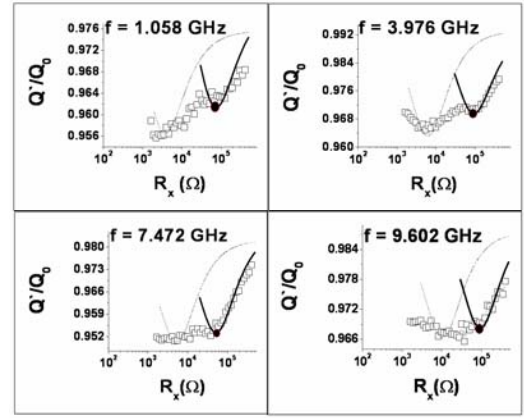

Fig 3: Symbols show the averaged $Q`/Q_0$ line cuts through Fig. 1(c) for four different frequencies. Also shown are the $Q`/Q_0$ fits to the data, where $Q_0$ is the quality factor of the resonator with no sample present. The fit parameters (see Table I) for the low-$R_x$ data (dashed lines) differ from the high-$R_x$ data (solid lines). The dots on the curve signify points where $\omega C_x(AR_x) = 1$.